\documentclass{article}%
\usepackage[top=3cm, bottom=3cm, left=3cm, right=3cm]{geometry}
\usepackage{amsmath}
\usepackage{amsfonts}
\usepackage{amssymb}
\usepackage{graphicx}%
\begin{document}

\title{Decays of open charmed mesons in the extended Linear Sigma Model  %
\thanks{Based on the poster preperation given by W. I. Eshraim at the 13th International Workshop on Meson Production, Properties and Interaction, 29th May - 3rd June 2014, KRAKÓW, POLAND. The poster was awarded with the special prize of the Poster Session. }}%

\author{Walaa I. Eshraim $^1$  and
      Francesco Giacosa $^{1,2}$\\\emph{$^1$ Institute for Theoretical Physics, Goethe University,}
\\\emph{Max-von-Laue-Str.\ 1, D--60438 Frankfurt am Main, Germany}\\\emph{ $^2$ Institute of Physics, Jan Kochanowski University, 25-406 Kielce, Poland}}
\maketitle

\begin{abstract}
We enlarge the so-called extended linear Sigma model
(eLSM) by including the charm quark according to the global
$U(4)_r\times U(4)_l$ chiral symmetry. In the eLSM, besides scalar
and pseudoscalar mesons, also vector and axial-vector mesons are
present. Almost all the parameters of the model were fixed in a
previous study of mesons below 2 GeV. In the extension to the
four-flavor case, only three additional parameters (all of them
related to the bare mass of the charm quark) appear. We
compute the (OZI-dominant) strong decays of open charmed mesons. The
results are compatible with the experimental data, although the theoretical  
uncertainties are still large. 
\end{abstract}

\section{Introduction}\label{intro} 

\indent Charm physics is an experimentally and theoretically active field of
hadronic physics \cite{brambilla}. The study of strong decays of the heavy mesons into light pseudoscalar
mesons is useful to classify the charmed states. In this paper we study the
OZI-dominant decays of open charmed mesons by using the extended
linear sigma model (eLSM).\newline
\indent The eLSM is a chiral model which describes successfully the vacuum
phenomenology of light mesons such as masses and decay widths  \cite%
{nf2D,Dick}. The eLSM Lagrangian is invariant under chiral symmetry $%
U(N_{f})_{R}\times U(N_{f})_{L}$ ($N_f=2$ in Ref.\cite{nf2D} and $N_f=3$ in Ref.\cite{Dick}) and  dilatation transformations. Chiral
symmetry is broken explicitly by non-vanishing quark masses and
spontaneously by a non-zero quark condensate in the QCD vacuum.%
\newline
\indent In this paper we extend the eLSM by including the charm degree of
freedom, thus working with four flavor ($N_{f}=4$) \cite{Proc1W, PapW}. The
masses of open and hidden charmed mesons have been calculated \cite%
{Proc1W, PapW} and turn out to be compatible with the experimental data \cite%
{PDG}. Here we then concentrate on the decay widths of open charmed
mesons \cite{PapW}.

\section{The eLSM and its implications}\label{sec-1} 

\indent In order to extend the eLSM to the case of four flavour, $N_{f}=4$, we introduce the (pseudo)scalar and
(axial-)vector meson fields in terms of $%
4\times 4$ (instead of $3\times 3$) matrices, which the charmed mesons appear in the
fourth row and column. The multiplet of (pseudo)scalar states is given by
\begin{equation}
\Phi =S+iP=\frac{1}{\sqrt{2}}\left( 
\begin{array}{cccc}
\frac{(\sigma _{N}+a_{0}^{0})+i(\eta _{N}+\pi ^{0})}{\sqrt{2}} & 
a_{0}^{+}+i\pi ^{+} & K_{0}^{\ast +}+iK^{+} & D_{0}^{\ast 0}+iD^{0} \\ 
a_{0}^{-}+i\pi ^{-} & \frac{(\sigma _{N}-a_{0}^{0})+i(\eta _{N}-\pi ^{0})}{%
\sqrt{2}} & K_{0}^{\ast 0}+iK^{0} & D_{0}^{\ast -}+iD^{-} \\ 
K_{0}^{\ast -}+iK^{-} & \overline{K}_{0}^{\ast 0}+i\overline{K}^{0} & \sigma
_{S}+i\eta _{S} & D_{S0}^{\ast -}+iD_{S}^{-} \\ 
\overline{D}_{0}^{\ast 0}+i\overline{D}^{0} & D_{0}^{\ast +}+iD^{+} & 
D_{S0}^{\ast +}+iD_{S}^{+} & \chi _{C0}+i\eta _{C}%
\end{array}%
\right) \,,  \label{Phi}
\end{equation}%
and the left-handed and right-handed matrices containing the vector fields and axial-vector fields are given by 
\begin{equation}
L^{\mu }(R^{\mu })=V^{\mu }\,\pm A^{\mu }\,=\frac{1}{\sqrt{2}}\left( 
\begin{array}{cccc}
\frac{\omega _{N}+\rho ^{0}}{\sqrt{2}}\pm \frac{f_{1N}+a_{1}^{0}}{\sqrt{2}}
& \rho ^{+}\pm a_{1}^{+} & K^{\ast +}\pm K_{1}^{+} & D^{\ast 0}\pm D_{1}^{0}
\\ 
\rho ^{-}\pm a_{1}^{-} & \frac{\omega _{N}-\rho ^{0}}{\sqrt{2}}\pm \frac{%
f_{1N}-a_{1}^{0}}{\sqrt{2}} & K^{\ast 0}\pm K_{1}^{0} & D^{\ast -}\pm
D_{1}^{-} \\ 
K^{\ast -}\pm K_{1}^{-} & \overline{K}^{\ast 0}\pm \overline{K}_{1}^{0} & 
\omega _{S}\pm f_{1S} & D_{S}^{\ast -}\pm D_{S1}^{-} \\ 
\overline{D}^{\ast 0}\pm \overline{D}_{1}^{0} & D^{\ast +}\pm D_{1}^{+} & 
D_{S}^{\ast +}\pm D_{S1}^{+} & J/\psi \pm \chi _{C1}%
\end{array}%
\right) ^{\mu }.  \label{4}
\end{equation}%

The charmed fields $D^{\ast 0},\,D^{\ast },\,D_{0}^{\ast
0},\,D_{0}^{\ast },\,D_{S0}^{\ast },\,D_{1},\,D_{S1},\,\chi _{c1},$
$\chi_{c0}$, and $J/\psi $ are assigned to the $q\overline{q}$ resonances $%
D^{\ast }(2007)^{0},\,D^{\ast }(2010)^{\pm },\,D_{0}^{\ast
}(2400)^{0},\,D_{0}^{\ast }(2400)^{\pm },\,D_{S0}^{\ast
}(2317),\,D_{1}(2420),$ $D_{S1}(2536)$ $\chi _{c1}(1P),\,\chi _{c0}(1P)$,
and $J/\psi (1S)$, respectively, see the details in Ref. \cite{PapW}. The
explicit form of the eLSM Lagrangian for $N_{f}=4$  is analogous to the case 
$N_{f}=3$ of Ref. \cite{Dick} (but has an additional term $-2\,\mathrm{Tr}%
[E\Phi ^{\dagger }\Phi]$):   
\begin{align}
\mathcal{L}& =\frac{1}{2}(\partial _{\mu }G)^{2}-\frac{1}{4}\frac{m_{G}^{2}}{%
\Lambda ^{2}}\left( G^{4}\,\log \frac{G}{\Lambda }-\frac{G^{4}}{4}\right) +%
\mathrm{Tr}[(D^{\mu }\Phi )^{\dagger }(D^{\mu }\Phi )]-m_{0}^{2}\left( \frac{%
G}{G_{0}}\right) ^{2}\mathrm{Tr}(\Phi ^{\dagger }\Phi )-\lambda _{1}[\mathrm{%
Tr}(\Phi ^{\dagger }\Phi )]^{2}  \nonumber \\
& -\lambda _{2}\mathrm{Tr}(\Phi ^{\dagger }\Phi )^{2}+\mathrm{Tr}[H(\Phi
+\Phi ^{\dagger })]+\mathrm{Tr}\left\{ \left[ \left( \frac{G}{G_{0}}\right)
^{2}\frac{m_{1}^{2}}{2}+\Delta \right] \left[ (L^{\mu })^{2}+(R^{\mu })^{2}%
\right] \right\} -\frac{1}{4}\mathrm{Tr}[(L^{\mu \nu })^{2}+(R^{\mu \nu
})^{2}]  \nonumber \\
& -2\,\mathrm{Tr}[E\Phi ^{\dagger }\Phi ]+c(\mathrm{det}\Phi -\mathrm{det}%
\Phi ^{\dagger })^{2}+i\frac{g_{2}}{2}\{\mathrm{Tr}(L_{\mu \nu }[L^{\mu
},L^{\nu }])+\mathrm{Tr}(R_{\mu \nu }[R^{\mu },R^{\nu }])\}  \nonumber \\
& +\frac{h_{1}}{2}\mathrm{Tr}(\Phi ^{\dagger }\Phi )\mathrm{Tr}[(L^{\mu
})^{2}+(R^{\mu })^{2}]+h_{2}\mathrm{Tr}[(\Phi R^{\mu })^{2}+(L^{\mu }\Phi
)^{2}]+2h_{3}\mathrm{Tr}(\Phi R_{\mu }\Phi ^{\dagger }L^{\mu })+\ldots \,,\,
\label{lag}
\end{align}%
where $G$ is the dilaton field, $D^{\mu }\Phi \equiv \partial ^{\mu }\Phi
-ig_{1}(L^{\mu }\Phi -\Phi R^{\mu })$,  $L^{\mu \nu }\equiv \partial ^{\mu
}L^{\nu }-\partial ^{\nu }L^{\mu }$, and $R^{\mu \nu }\equiv \partial ^{\mu
}R^{\nu }-\partial ^{\nu }R^{\mu }$. The terms $\mathrm{Tr}[H(\Phi +\Phi
^{\dagger })]$ with $H=1/2\,\text{diag}\{h_{0N},\,h_{0N},\,\sqrt{2}h_{0S},\,\sqrt{2}%
h_{0C}\}$, $-2\,\mathrm{Tr}%
[E\Phi ^{\dagger }\Phi]$ with $E=\text{diag}\{\varepsilon _{N},\,\varepsilon
_{N},\,\varepsilon _{S},\,\varepsilon _{C}\},\,\varepsilon _{i}\propto
m_{i}^{2},\,\varepsilon _{N}=\varepsilon _{S}=0$, and $\mathrm{Tr}\left[
\Delta (L^{\mu }{}^{2}+R^{\mu }{}^{2})\right]$ with $\delta =\text{diag}\{\delta
_{N},\,\delta _{N},\,\delta _{S},\,\delta _{C}\},\,\delta _{i}\sim
m_{i}^{2},\,\delta _{N}=\delta _{S}=0$, break chiral symmetry due to nonzero
quark masses and are especially important for mesons containing the charm
quark. When $m_{0}^{2}<0$ spontaneous symmetry breaking occurs and the
fields $\sigma _{N}$, $\sigma _{S}$, and $\chi _{C0}$ condense \cite%
{Dick,Proc1W, PapW}. Most of the parameters of the model were already fixed
in the three-flavor study of Ref. \cite{Dick}. Only three new parameters
appear and all of them are related to the bare mass of the charm quark. They
were determined in Ref. \cite{PapW} through a fit to the masses of charmed
mesons. As an outcome, the charm-anticharm condensate is sizable, $%
\left\langle \chi _{C0}\right\rangle =178\pm 28$  MeV. 

\begin{table}[tbp]
\caption{Decay widths of open charmed mesons}
\label{tab-1}\centering 
\begin{tabular}{lll}
\hline
Decay Channel & Theoretical results [MeV] & Experimental results [MeV] (from \cite{PDG})\\ 
\hline
$D_{0}^{\ast}(2400)^{0}\rightarrow D\pi$ & $139_{-114}^{+243}$ & $%
D^{+}\pi^{-}$ seen; full width $\Gamma=267\pm 40$ \\ 
$D_{0}^{\ast}(2400)^{+}\rightarrow D\pi$ & $51_{-51}^{+182}$ & $D^{+}\pi^{0}$
seen; full width: $\Gamma=283\pm24\pm 34$ \\ 
$D^{\ast}(2007)^{0}\rightarrow D^{0}\pi^{0}$ & $0.025\pm0.003$ & seen; $<1.3$
\\ 
$D^{\ast}(2007)^{0}\rightarrow D^{+}\pi^{-}$ & $0$ & not seen \\ 
$D^{\ast}(2010)^{+}\rightarrow D^{+}\pi^{0}$ & $0.018_{-0.003}^{+0.002}$ & $%
0.029\pm0.008$ \\ 
$D^{\ast}(2010)^{+}\rightarrow D^{0}\pi^{+}$ & $0.038_{-0.004}^{+0.005}$ & $%
0.065\pm0.017$ \\ 
$D_{1}(2420)^{0}\rightarrow D^{\ast}\pi$ & $65_{-37}^{+51}$ & $%
D^{\ast+}\pi^{-}$ seen; full width: $\Gamma=27.4\pm 2.5$ \\ 
$D_{1}(2420)^{0}\rightarrow D^{0}\pi\pi$ & $0.59\pm0.02$ & seen \\ 
$D_{1}(2420)^{0}\rightarrow D^{+}\pi^{-}\pi^{0}$ & $0.21_{-0.015}^{+0.01}$ & 
- \\ 
$D_{1}(2420)^{0}\rightarrow D^{+}\pi^{-}$ & $0$ & not seen; $%
\Gamma(D^{+}\pi^{-})/\Gamma(D^{\ast+}\pi^{-})<0.24$ \\ 
$D_{1}(2420)^{+}\rightarrow D^{\ast}\pi$ & $65_{-36}^{+51}$ & $%
D^{\ast0}\pi^{+}$ seen; full width: $\Gamma=25\pm 6$ \\ 
$D_{1}(2420)^{+}\rightarrow D^{+}\pi\pi$ & $0.56\pm0.02$ & seen \\ 
$D_{1}(2420)^{+}\rightarrow D^{0}\pi^{0}\pi^{+}$ & $0.22\pm0.01$ & - \\ 
$D_{1}(2420)^{+}\rightarrow D^{0}\pi^{+}$ & $0$ & not seen; $%
\Gamma(D^{0}\pi^{+})/\Gamma(D^{\ast0}\pi^{+})<0.18$ \\ \hline
\end{tabular}
\end{table}

\section{Results and Conclusion}

We present the results of the decay widths of open charmed mesons in Table
1. Although the theoretical errors are in some channels very large, the
results are compatible with the experiment \cite{PDG}. Moreover, the decay
of the vector and axial-vector chiral partners $D^{\ast }(2010)$ and $%
D_{1}(2420)$ are well described. This fact shows that chiral symmetry is
still important for charmed mesons. Studies in progress are the calculation
of the decay widths of the charmonia and the mixing of axial-vector and
pseudovector charmed states. 

\section*{Acknowledgments}

The authors thank 
D.\ Parganlija and D. H.\ Rischke for
useful discussions. W.I.E.\ acknowledges financial support from DAAD.


\begin{thebibliography}{00}

\bibitem{brambilla} 
N.~Brambilla, A.~Pineda, J.~Soto and A.~Vairo,
Rev.\ Mod.\ Phys.\  {\bf 77} (2005) 1423.

\bibitem{nf2D} 
  D.~Parganlija, F.~Giacosa and D.~H.~Rischke,
  Phys.\ Rev.\ D {\bf 82}, 054024 (2010).
  
\bibitem{Dick} 
  D.~Parganlija, P.~Kovacs, G.~Wolf, F.~Giacosa and D.~H.~Rischke,
  Phys.\ Rev.\ D {\bf 87}, 014011 (2013).
  
\bibitem{Proc1W}
  W.~I.~Eshraim,
  PoS QCD {\bf -TNT-III}, 049 (2013)
  [arXiv:1401.3260 [hep-ph]].
  
\bibitem{PapW}
  W.~I.~Eshraim, F.~Giacosa and D.~H.~Rischke,
  arXiv:1405.5861 [hep-ph].
  
\bibitem {PDG} J.~Beringer \textit{et al}. (particle Data Group), Phys. Rev. D
\textbf{86}, 010001 (2012).

\end{thebibliography}
\end{document}